\documentclass{style_arxiv}
\usepackage{amssymb}
\usepackage{mathtools}
\usepackage{siunitx}

\def\Journal#1#2#3#4{{#1} {\bf #2}, #3 (#4)}

\def\PLB{{\em Phys. Lett.}  B}
\def\PRL{\em Phys. Rev. Lett.}
\def\PRD{{\em Phys. Rev.} D}
\def\PRC{{\em Phys. Rev.} C}

\def\JHEP{{\em JHEP}}

\def\barnue{\bar{\nu}_e}
\def\nue{{\nu}_e}
\def\be{\begin{equation}}
\def\ee{\end{equation}}
\def\bea{\begin{eqnarray}}
\def\eea{\end{eqnarray}}

\begin{document}

\headline{Proceedings for the 29th Rencontres de Blois 2017 on Particle Physics and Cosmology}

\vspace*{4cm}
\title{LIGHT STERILE NEUTRINO SEARCHES}

\author{ J. HASER }

\address{Max-Planck-Institut f\"{u}r Kernphysik, Saupfercheckweg 1,\\
69117 Heidelberg, Germany}

\maketitle\abstracts{Several anomalies are discussed in neutrino physics, linked to the possible existence of light sterile neutrinos. These sterile neutrinos would not interact weakly, but they could leave an imprint in various measurements, such as neutrino oscillation experiments or precise measurements of beta decay spectra. Global analyses of neutrino oscillation data suggest sterile neutrinos with a mass in the eV range, but reveal at the same time tensions with the observed neutrino anomalies. In the next years several experiments will test the sterile neutrino hypothesis with masses in the eV range with promising sensitivities.}

\vspace*{2cm}

\section{Introduction}

The neutrino, probably the most peculiar particle of today's Standard Model of particle physics, has always been good for a surprise. Its history started in 1934, back then as a hypothetical particle postulated by Wolfgang Pauli, to save the laws of energy and momentum conservation. After several failed attempts to prove its existence, it was finally detected by F.~Reines and C.~Cowan in the '50s near a nuclear reactor. Since neutrinos, oddly enough, are only observed as left-handed particles and antineutrinos as right-handed, they were assumed to be massless. Another surprising property, however, proved this assumption wrong and can be taken as the first indication of physics beyond the standard model: neutrinos change their flavour during flight, implying that their flavour eigenstates are superpositions of three mass eigenstates of different masses. The unambiguous observation of neutrino flavour oscillations, and hence the existence of neutrino mass, gained A.~McDonald and T.~Kajita the Nobel Prize in 2015.\\
Three anomalies in neutrino physics, which are possibly linked to each other, have yet to be solved. All three could be --more or less-- explained by the existence of light sterile neutrinos with a mass in the eV range. Various experiments try and will try to test the favoured parameter space in the upcoming years, many of them with different neutrino sources and detection concepts.

\newpage

\section{Current anomalies in neutrino physics}

The three anomalies mentioned above were observed in experiments with different sources, detection channels and detector principles. They either observed a deficit in measured neutrino rate or detected an excess of electron (anti)neutrinos.
\\\\
In the 1990s the solar neutrino experiments GALLEX~\cite{1999gallex} and SAGE~\cite{2009sage} performed measurements with ${}^{51}$Cr and ${}^{37}$Ar sources of $\sim\,1$MCi activity, both $\beta^+$-emitters and hence electron neutrino sources. The goal of these calibration measurements, in which the sources were deployed in or near the detectors, was to demonstrate that the detection technology works as expected. The measurement helped to assure the community that the observed deficit in solar neutrino flux is not linked to unknown inefficiencies in the detection principle. However, both experiments could not meet the expected neutrino rate with their measurements at a significance~\cite{2011giunti} of about $3\,\sigma$: The two ${}^{51}$Cr measurements in GALLEX and the ${}^{51}$Cr and ${}^{37}$Ar measurements in SAGE yield an average observed-to-predicted rate ratio of $\bar{R} = 0.85 \pm 0.05$.
\\\\
The experiments LSND and MiniBooNE were two accelerator produced neutrino experiments, which observed an excess of electron antineutrinos in a beam of muon antineutrinos. LSND used a stopped pion beam of muon antineutrinos with energies in the range of 20 to 55\,MeV as source. The detector was placed at a distance of 30 metres from the beam target. The excess in electron antineutrinos observed by the LSND experiment~\cite{1997lsnd}, which took data from 1993 to 1998, were significant at more than $3\,\sigma$ and could not be fully excluded by the KARMEN experiment~\cite{2002karmen} in 2002. The succeeding experiment, MiniBooNE, was a 800 ton detector, filled with mineral oil. It was placed at about 500 metres from the point at which the neutrino beam is produced with an average energy of about 500\,MeV. Having first announced an exclusion~\cite{2007mboone} of the LSND findings in 2007 at 98\,\% C.L. from an analysis of a neutrino beam, the MiniBooNE collaboration published in 2010 the result from antineutrino data, which claimed an $\barnue$ event excess~\cite{2010mboone} in agreement with LSND. Finally, in 2013, a combined analysis~\cite{2013mboone} of the collected $\nue$ and $\barnue$ data was released, reporting an event excess at $3.8\,\sigma$ significance. In this new data set, both, the neutrino and antineutrino measurements, showed an excess in electron flavour events at low energies (below 0.5\,GeV). These new results were taken as hint for the existence of sterile neutrinos with $\sim\,1\,$eV mass, but the neutrino and antineutrino data were not in perfect accordance. Only if neutrinos were allowed to behave differently than antineutrinos, the results would agree, requiring a CP violating phase and hence more than one sterile neutrino.
\\\\
The third anomaly was observed by reactor neutrino experiments, after a re-evaluation of the predicted reactor spectra. In a nuclear reactor the main neutrino flux ($0 < E_{\bar{\nu}_e} < 10\,$MeV) is produced by the fission fragments of the four actinides ${}^{235}$U, ${}^{239}$Pu, ${}^{238}$U and ${}^{241}$Pu. High-precision measurements of the electron spectra of fissioning ${}^{235}$U, ${}^{239}$Pu and ${}^{241}$Pu were collected. From these, the neutrino spectra can be deduced following the laws of energy conservation and under application of a set of energy dependent corrections. Until 2011, a set of reference spectra computed by Schreckenbach et al.~were used (for ${}^{238}$U the spectrum was built based on nuclear data bases). The error on these spectra, however, were with 3\,\% on average a significant uncertainty in the analysis of the upcoming generation of neutrino experiments~\cite{2012dc,2012db,2012reno}, trying to measure the smallest neutrino mixing angle $\theta_{13}$. This triggered a re-evaluation of the neutrino spectra by two independent groups\cite{2011mueller,2011huber}, using the very same beta-spectra Schreckenbach et al.~had measured and used, but this time with more input from updated nuclear data bases and refined approaches in the application of the correction factors. Both groups could not yield a reduction in the uncertainties with their analyses, but both found a $\sim\,4\,\%$ increase in the absolute neutrino flux. Together with an increase of about 1.5\,\% in neutrino detection cross section, which was linked to a change of the measured neutron lifetime, the predicted neutrino flux detected by reactor experiments went up by almost 6\,\%. A re-analysis of 19 reactor experiments with baselines of 100 metres and less, revealed a deficit in the measured neutrino flux~\cite{2011mention,2013lasserre}: the observed-to-predicted ratio of the absolute neutrino rate was found to be $R = 0.936 \pm 0.024$. This deficit of $2.7\,\sigma$ significance is known as \textit{Reactor Antineutrino Anomaly} (RAA).
\\
Another anomaly was observed in the spectral shape of neutrinos produced at nuclear reactors~\cite{2014dc,2014reno,2015db,2017neos}. At $E_{\bar{\nu}_e} \sim 5\,$MeV a deviation is found compared to the shape of the reference spectra with $>3\,\sigma$ significance. This anomaly is seen by detectors with different baselines to the reactor core and hence not considered as possible hint of a light sterile neutrino signal.

\section{Light sterile neutrino hypothesis}

The known neutrino mixing parameters, in this case the differences of the squared masses $\Delta m^2$, induce neutrino flavour oscillations at baseline-to-energy ratios $L/E \gtrsim 1\,\mathrm{km}/\mathrm{MeV}$. All three anomalous observations, however, were made at $L/E \approx 1\,\mathrm{m}/\mathrm{MeV}$. Nevertheless, they could be explained by neutrino oscillations with $\Delta m^2_\mathrm{new} \gtrsim 1\,\mathrm{eV}^2$, i.e.~neutrino oscillations from active to sterile state.
\\
In a minimal extension of the standard model, a fourth neutrino mass eigenstate is added with a mass in the eV range, in order to explain the experimental data. Accordingly, the neutrino mixing matrix $U_\mathrm{PMNS}$ is extended to a $4\times 4$ matrix, and introduces at the same time a fourth new ``flavour'' state $|\nu_s\rangle$:
\begin{eqnarray} \label{eqU}
	\left( \begin{array}{c} \nu_{e}  \\   \nu_{\mu}    \\    \nu_{\tau}  \\    \nu_{s}   \\ \end{array} \right)  =  \left( \begin{array}{cccc} U_{e1}  &   U_{e2}   &    U_{e3}  &    U_{e4}   \\   U_{\mu 1}   &   U_{\mu 2} &  U_{\mu 3}  &  U_{\mu 4} \\  U_{\tau 1}   &  U_{\tau 2} &  U_{\tau 3} &  U_{\tau 4} \\ U_{s1}  &   U_{s2}   &    U_{s3}  &    U_{s4}   \\ \end{array} \right)  \left( \begin{array}{c} \nu_{1}  \\   \nu_{2}    \\    \nu_{3}  \\    \nu_{4}   \\ \end{array} \right) \,.\\ \nonumber
\end{eqnarray}
In this ``3+1 model'' --with three active and one sterile neutrino-- the fourth mass eigenstate is taken to be mostly sterile with $|U_{s4}|^2 \approx 1$ and $|U_{e4}|^2,|U_{\mu4}|^2,|U_{\tau4}|^2 \ll 1$.
Since the mass of the added mass eigenstate is $\mathcal{O}\sim\,\mathrm{eV}$, the new ``flavour'' state cannot be weakly interacting in accordance with the Z boson width measurements at LEP~\cite{2006lep}, the new neutrino must be \textit{sterile}.
\\
In the extended ``3+1 model'' the flavour oscillation probabilities change. Using the three anomalies as input, global fits can be performed and yield contours enclosing the favoured parameters describing the new oscillation. When the RAA and the GALLEX/SAGE observation is combined with all other experimental data on electron flavour neutrino experiments (which did not see an anomalous signal), the global analysis~\cite{2013kopp} yields for the new oscillation the best fit parameters $|\Delta m_\mathrm{new}^2| = 1.8\,\mathrm{eV}^2$ and $\sin^22\theta_\mathrm{new} = 0.09$. This means, the fourth mass eigenstate is favoured to have a mass of few eV, and the maximal observable deficit of electron flavour due to oscillation is 9\,\%. Adding the anomaly of appearing electron flavour (except for part of the MiniBooNE data at low energies $E<475\,$MeV) and all other available neutrino oscillation data as of 2013, the favoured oscillation parameters become
$|\Delta m_\mathrm{new}^2| = 1.6\,\mathrm{eV}^2$ and $\sin^22\theta_\mathrm{new} = 0.11$. In this analysis~\cite{2013giunti} the no-oscillation hypothesis was excluded at $6\,\sigma$.

\section{Sterile neutrino signal and searches}

The new suggested neutrino does not interact weakly and can therefore not be directly detected. Still, there are various possibilities to observe it. The key to prove its existence is the detection of a change in expectation values due to a fourth mass eigenstate.
This can either be realised in an oscillation signature, or in a change of the beta spectrum shape measured with very high precision. Both can be unambiguous observations of light sterile neutrinos, especially if an oscillation experiment sees a variation of the survival (appearance) probability with respect to neutrino energy and detection baseline.
Some of the already existing neutrino detectors can be used to test the sterile neutrino hypothesis, others are built or have been built in the past few years for this exact purpose.
\\\\
There are two standard ways to detect neutrino oscillations, i.e.~neutrino flavour transitions: \textit{appearance experiments}, trying to detect neutrinos of a certain flavour in a flux of different initial flavour, and \textit{disappearance experiments}, which search for a deficit of neutrino flux. No matter which of the two approaches is followed, the baseline-to-energy ratio has to be of the order $\sim\,1\mathrm{m}/\mathrm{MeV}$ to be sensitive to a clear signature of a sterile neutrino with $\sim\,1$eV mass.
\\
Concerning the logistics and realization of such a sterile search experiment, on can either bring a neutrino source to a large detector, or build a detector in the vicinity of an existing neutrino source. For each of the approaches an uncertainty on the few percent level on the signal is required, which can be achieved by strong sources, large target masses, high signal-to-background ratios and good event identification capabilities. Furthermore, a compact source is of advantage as much as good vertex and energy resolution of the detector response, to not smear the oscillation signature. Information on the absolute normalization of the initial neutrino flux can further improve the experiment's sensitivity, especially with respect to sterile neutrino masses larger than few eV.

\subsubsection{Tritium beta-decay spectrum and KATRIN}

The KATRIN experiment (Karlsruhe, Germany) is, unlike the other experiments discussed in this manuscript, not a neutrino flavour oscillation experiment. The detector is a large $\beta$-spectrometer, measuring the electron spectrum from tritium decay near the Q-value $E_0 ({}^{3}_{1}\mathrm{T})$ with very high precision. In this way, it tries to determine the mass of the electron antineutrino. The electron flavour eigenstate is a superposition of the mass eigenstates $i$, therefore the tritium spectrum is a sum of $\beta$-spectra with different endpoints $E_0 ({}^{3}_{1}\mathrm{T}) - m_{\nu,i}$. Since the mass splittings between the different eigenstates are too small to be experimentally resolved, KATRIN measures an ``effective electron neutrino mass'' $m_{\barnue}$ near the Q-value. In a similar way, the experiment could search for additional, heavier mass eigenstates~\cite{2015mertensa,2015mertensb}. The measured $\beta$-spectrum $\frac{d\Gamma}{dE}$ is given by
\begin{equation}
\frac{d\Gamma}{dE} = \cos^2\theta \frac{d\Gamma}{dE}(m_\mathrm{light}) + \sin^2\theta \frac{d\Gamma}{dE}(m_\mathrm{heavy})\,,
\end{equation}
with an active-to-sterile mixing amplitude $\sin^2\theta$. An additional mass eigenstate would lead to a spectral distortion and would leave a kink in the spectrum  at $E_0 ({}^{3}_{1}\mathrm{T})- m_\mathrm{heavy}$. The projected sensitivity of KATRIN improves with increasing sterile neutrino mass and has access to parameter space regions of masses larger than few eV and $\sin^2(2\theta) < 0.1$ (95\,\% C.L.), where other experiments already lose sensitivity. Furthermore, with an upgrade of KATRIN to yield higher event statistics, the experiments could be sensitive to keV-scale sterile neutrinos, which are candidates for warm dark matter.

\subsubsection{Atmospheric neutrinos and IceCube}

The IceCube neutrino telescope is located at the South Pole, Antarctica, and detects neutrino interactions in the Antarctic ice with an array of photosensors spanning a volume of about one cubic metre. Via the detection of atmospheric neutrinos, the experiment searches in the pattern of muon flavour disappearance for an imprint caused by sterile neutrinos. For this analysis~\cite{2016aartsen} it detects neutrinos with ``upward'' pointing travel paths, i.e.~neutrinos which have traversed the Earth and travelled distances of $L < 1.2 \cdot 10^4\,$km. As the energy of the neutrinos is in the range of $320\,\mathrm{GeV} < E < 20\,\mathrm{TeV}$, the analysis is sensitive to $L/E = [0.01, 10]\,\mathrm{m}/\mathrm{MeV}$ and hence sterile neutrinos of eV masses. Matter effects of neutrinos passing the Earth's mantle and core would cause resonant active-sterile oscillations, which would amplify a sterile neutrino signature. The experiment tested the $\Delta m^2$ range from 0.1 to $10\,\mathrm{eV}^2$ not finding a hint for an unexpected $\nu_\mu$ or $\bar{\nu}_\mu$ disappearance. In the mass region around $0.3\,\mathrm{eV}^2$ IceCube could establish a limit on $\sin^2(2\theta_{24}) \leq 0.05$ at 99\,\% C.L.
\\\\
With the DeepCore detector extension of the IceCube observatory by additional photosensors, atmospheric neutrinos of energies below $100\,$GeV could be detected. With three years of data~\cite{2017aartsen}, the limits $|U_{\mu 4}|^2 < 0.11$ and $|U_{\tau 4}|^2 < 0.15$ were set, both at 90\,\% C.L.

\subsubsection{Accelerator based searches}

Sterile neutrino searches at particle accelerators offer the possibility to neutrino flavour disappearance as well as appearance measurements. One of them, the Short-Baseline Neutrino (SBN) physics program~\cite{2015antonello}, uses an accelerator decay in flight beam, from Booster Neutrino Beam at Fermilab to three LAr-TPC detectors. The closest detector SBND (Short Baseline Near Detector) will be located at 110 metres distance and have an active mass of 112 tons. At 470 metres MicroBooNE will be placed, with a fiducial volume mass of 89 tons. The third LAr-TPC, ICARUS, will have a baseline of 600 metres and a mass of 476 tons. The physics program suggests measurements of $\nu_\mu$ as well as $\nu_e$ disappearance in neutrino and antineutrino mode, and tests of electron flavour appearance in a muon neutrino beam. With $10^{20}$ to $10^{21}$ protons on target, the SBN program could test the LSND 99\,\% C.L.~region at the  $5\,\sigma$ level.

\subsubsection{${}^{144}$Ce source experiment SOX}

A sort of refined repetition of the Gallium calibration measurements of GALLEX and SAGE is realised by the SOX experiment~\cite{2013bellini,2015vivier}. The large solar neutrino detector Borexino is used for this purpose, which is a 270 ton liquid scintillator detector. A reactor produced 100\,kCi (3.7\,PBq) ${}^{144}$Ce source will be placed below the detector, 8.5 metres from the detector centre. The emitted electron antineutrinos are then detected via a distinct coincidence signal from the \textit{inverse beta decay} reaction, which has an energy threshold of $1.8\,$MeV. Such high activities together with an endpoint above the detection threshold can be achieved by means of a two step decay, as in the case of ${}^{144}$Ce:
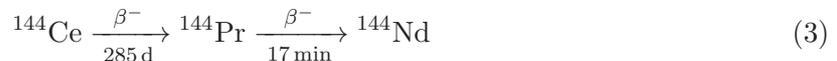
\begin{equation}
{}^{144}\mathrm{Ce} \xrightarrow[\;285\,\mathrm{d}\;]{\beta^-} {}^{144}\mathrm{Pr} \xrightarrow[\;17\,\mathrm{min}\;]{\beta^-} {}^{144}\mathrm{Nd}
\end{equation}
The ${}^{144}$Ce $\beta^-$-decays with a Q-value of 318\,keV, while the short-lived ${}^{144}$Pr has a Q-value of about 3\,MeV. The two-fold coincidence of the detection reaction leads to negligibly small background rates, the position resolution is about 15\,cm and the energy resolution amounts to 5\,\% at 1\,MeV visible energy. Knowing the source activity with about 1\,\% precision and with an experimental error of 2\,\%, uncorrelated between the energy and space bins, a sensitivity of $\sin^2(2\theta) \lesssim 0.06$ can be reached for $\Delta m^2 \sim 1\,\mathrm{eV}^2$ at 95\,\% C.L. after 1.5 years of data taking.

\subsubsection{Reactor antineutrino experiments}

A large number of reactor neutrino experiments have been performed in the past decades, the most popular ones measured the neutrino oscillation parameters of the three known active neutrinos. Among these are experiments which could not only measure the neutrino flux but also the energy spectrum and therefore perform sterile neutrino searches (e.g.~Double Chooz, Daya Bay, Bugey-3). The baselines, however, were too large to be sensitive to $\Delta m^2 \gtrsim 1\,\mathrm{eV}^2$. A new generation of experiments is hence built at shorter baselines of about 20 metres and less. Ideally an oscillation signal is observed as distortion of the measured neutrino spectrum, which changes with respect to the measurement baseline. In order to observe such a dependence and gain sensitivity, most of these new detectors are segmented, movable or both at the same time.
\\\\
Nuclear reactors are strong neutrino sources, emitting about $2 \cdot 10^{17}$ electron antineutrinos per MW thermal power and second. The $\barnue$ are then detected via \textit{inverse beta decay} (IBD, $\barnue + p \rightarrow e^+ + n$) on a hydrogen nucleus, a reaction only sensitive to electron flavour, since reactor neutrinos have $\lesssim10\,$MeV energies. Two particles are produced in this weak interaction: a positron, which carries the energy information of the neutrino, and a neutron with $15\,$keV kinetic energy on average. Organic scintillators as detector material, liquid or solid, have proven to be beneficial for several reasons. Scintillating materials allow to detect the positron ionization signal and reconstruct the neutrino spectrum. Furthermore, organic scintillators consist of hydrocarbons, providing hydrogen atoms and thus the neutrino target. A reliable position resolution is obtained via segmentation of the detectors, a high segmentation also enhances the background rejection capabilities. The detector material is loaded or coated with isotopes of high neutron capture cross section, to use the neutron capture as tag for the neutrino interaction in order to suppress background events. Neutron captures on Gd yields a multi-gamma cascade with a total energy of $8\,$MeV, well above radioactive events of the natural decay chains. Gd-loaded liquid scintillators represent a mature technology, widely used in neutrino physics, for which long term stability has been proven both with respect to the chemical properties as well as physics performance. The use of ${}^{6}$Li for neutron capture can be advantageous, since the reaction produces a triton and an alpha particle, both particles with high $dE/dx$, which yields a distinct scintillation pulse shape. Moreover, the short range of both particles leads to a good energy containment.
\\\\
Table~\ref{table1} summarises the key characteristics of a selection of current very short baseline projects located at nuclear reactors. In the following paragraph three of the projects --NEOS, DANSS and Stereo-- are discussed in more detail.
\begin{table*}[b]
	\centering
	\caption{Selection of reactor neutrino experiments. Listed are the detector technology (PS: plastic scintillator, LS: liquid scintillator), the target mass $m_\mathrm{t}$, the thermal power of the reactor $P_\mathrm{th}$, the reactor to detector baseline $L$ and the signal-to-background ratio S/B. $R_\nu$ is the measured (or expected) neutrino rate at reactor on and shortest baseline. The photon statistical part of the energy resolution $\sigma_{E,\mathrm{Ph}}/E$ is given at \SI{1}{MeV} visible energy.}
	\vspace{1ex}
	\label{table1}
	\begin{tabular}{|l| >{\centering}p{1.8cm}<{\centering} >{\centering}p{1.1cm}<{\centering} p{1.5cm}<{\centering} p{1.6cm}<{\centering} p{1.5cm}<{\centering} p{1cm}<{\centering} p{1.7cm}<{\centering}|}
		\hline
		experiment	& technology &  $m_\mathrm{t}$ [t]  &  $P_\mathrm{th}$ [$\mathrm{MW}$]  &  $L$ [m]  &  $R_\nu$ [$\mathrm{day}^{-1}$]  &  S/B  &  $\sigma_{E,\mathrm{Ph}}/E$ \\
		\hline
		DANSS~\cite{2016danss}	& Gd-PS	& 0.9	& 3000	& 10.7-12.7 & 5000 	& 20	& 0.18  \\
		NEOS~\cite{2017neos}		& Gd-LS & 1	 & 2800	& 24  	& 1976 	& 22	&  0.05  \\
		Neutrino-4~\cite{2012serebrov,2013serebrov}	& Gd-LS & 1.4 & 90 	& 6-12	& 1800	& $\gtrsim1$	& -  \\
		Stereo~\cite{2015stereo} & Gd-LS	& 1.8	 & 57.8	  & 9-11	& 300	& $\sim1$	& 0.05  \\
		SoLid~\cite{2015solid,2017solid} 	& ${}^{6}$Li-PS	& 1.6 & 60-80	& 6-8	& 1200 	& $\sim1$	& 0.14	 \\
		Prospect~\cite{2016ashenfelter}	& ${}^{6}$Li-LS	& 3 & 85	& 7-12 	& 660	& 3	& 0.045	 \\
		\hline
	\end{tabular}
\end{table*}
\\
NEOS~\cite{2017neos} is a non-segmented 1 ton detector, deployed at 23 metres distance from one of the six reactor cores of the Hanbit power plant (Korea), where also the RENO experiment is located. The detector measured about 2000 IBD events per day with a remarkable signal-to-background ratio of 20. It took data for eight months. In the shape-only analysis, for which NEOS compared their measured spectrum to the Daya Bay spectrum as reference, they found no strong evidence to favour a 3+1 scenario over a model with only three neutrinos. In the $\Delta m^2_\mathrm{new}$ range from 0.2 to $2.3\,\mathrm{eV}^2$ they could limit $\sin^2(2\theta_\mathrm{new})$ to be below 0.1 at a confidence level larger than 90\,\%. The minimal $\chi^2$ value was found for the parameter set $(\sin^2(2\theta_\mathrm{new}), \Delta m^2_\mathrm{new}) = (0.05, 1.73\,\mathrm{eV}^2)$.
\\
The DANSS~\cite{2016danss} experiment was built at the Kalinin Nuclear Power Plant in Russia. It consists of stacked strips of composite Gd-plastic scintillator, forming 10 modules of 20\,cm width and 1\,m length, which cross each other. The detector is placed below a $3\,\mathrm{GW}_\mathrm{th}$ reactor and measures at three different heights from 10.7 to 12.7 metres about 5000 neutrino events per day. Their background contribution is 5\,\%, equivalent to a signal-to-background ratio of 20. From an analysis~\cite{2017danss} comparing a measurement of the neutrino spectrum in the upward position with a measurement in the downward position, the experiment could not find a hint for active-to-sterile oscillations. In the $\Delta m^2_\mathrm{new}$ range from 0.4 to $4\,\mathrm{eV}^2$ they could limit $\sin^2(2\theta_\mathrm{new})$ to be less than 0.2 at a confidence level larger than 95\,\%.
\\\\
The Stereo~\cite{2015stereo} project is sited at the research reactor of the Institute Laue-Langevin in Grenoble (France) and takes data since November 2016. Unlike the other two experiments, it measures antineutrinos from a core fueled with highly enriched uranium. The $2\,\mathrm{m}^3$ liquid scintillator detector is lengthwise divided in six separate cells, allowing for a multi-baseline measurement of the neutrino spectrum. An active-to-sterile oscillation would distort the neutrino spectrum of each cell differently. Stereo will be able to test the 99\,\% C.L.~region of the RAA at 95\,\% C.L.~with 300 live-days of data.

\section{Global Analyses}

As discussed above, a number of anomalous observations have been found in neutrino oscillation experiments, both in electron flavour disappearance (RAA, GALLEX, SAGE) and appearance (LSND, MiniBooNE). On the other hand, other experiments have not observed any evidence of non-standard oscillations, among them all measurements of muon flavour disappearance. The link between the different experimental results can be understood by looking at the parametrisation of neutrino oscillations. The survival and appearance probabilities  $P$ are often expressed in the 3+1 model short baseline approximation (with $L/E \sim 1\mathrm{m}/\mathrm{MeV}$) in the following way:
\begin{equation} \label{probee}
P(\nu_{e} \rightarrow \nu_{e}) \approx 1- \sin^2(2\theta_{ee}) \sin^2(1.27 \Delta m^2_{41} L/E)\,,
\end{equation}
\begin{equation} \label{probemu}
P(\nu_{\mu} \rightarrow \nu_{e}) \approx \sin^2(2\theta_{e\mu}) \sin^2(1.27 \Delta m^2_{41} L/E)\,,
\end{equation}
\begin{equation} \label{probemumu}
P(\nu_{\mu} \rightarrow \nu_{\mu}) \approx 1- \sin^2(2\theta_{\mu\mu}) \sin^2(1.27 \Delta m^2_{41} L/E)\,.
\end{equation}
Here, the $\sin^2(2\theta_{\alpha\alpha})$ (with $\alpha = e,\mu$) are effective flavour disappearance amplitudes. Likewise $\sin^2(2\theta_{e \mu})$ denotes the electron flavour appearance amplitude. The effective oscillation angles can be written in terms of the neutrino mixing matrix in Eq.~(\ref{eqU}), we then yield
\begin{equation} \label{anglee}
\sin^2(2\theta_{ee}) =4 |U_{e4}|^2 (1-|U_{e4}|^2)\,,
\end{equation}
\begin{equation} \label{angleemu}
\sin^2(2\theta_{e\mu}) = 4 |U_{e4}|^2 |U_{\mu 4}|^2\,,
\end{equation}
\begin{equation} \label{anglemu}
\sin^2(2\theta_{\mu \mu}) = 4 |U_{\mu 4}|^2 (1-|U_{\mu 4}|^2)\,.
\end{equation}
For a very common parametrization of neutrino mixing~\cite{2013kopp} used by global analyses we furthermore find $\theta_{ee} = \theta_{14}$ and the dependencies $|U_{e4}|=\sin\theta_{14}$ and $|U_{\mu 4}| = \cos\theta_{14} \sin\theta_{24}$. Comparing now Eq.~(\ref{probee})-(\ref{anglemu}) we see that an oscillation signal at very short baselines in the electron disappearance channel does not necessarily affect the muon disappearance channel ($\theta_{24}$ and consequently $|U_{\mu 4}|^2$ could be zero). A signal of electron flavour appearance, however, implies that both $|U_{e4}|^2$ and $|U_{\mu 4}|^2$ are non-zero, each supposing neutrino flavour disappearance of electron and muon flavour. Hence, a tension between the different data sets is found, especially when the stringent limits from MINOS and IceCube on a non-observation of muon flavour are included in global fits. The IceCube collaboration published a result~\cite{2016aartsen} in 2016 in which the LSND and MiniBooNE observation is excluded at $\sim\,99\,\%$ C.L. for the global best fit value~\cite{2013kopp,2013conrad} of $|U_{e4}|^2$.
\\
Most recent analyses~\cite{2017gariazzo} --which include also the NEOS, MINOS IceCube data-- find remaining islands in the ($\sin^2(2\theta)$, $\Delta m_{41}^2$) parameter space, narrow with respect to $\Delta m_{41}^2$. The best-fit point is found for $(\sin^2(2\theta_{14}), \Delta m_{41}^2) = (0.079, 1.7\,\mathrm{eV}^2)$. The appearance-disappearance tension discussed above can only be handled by neglecting the MiniBooNE data below 475\,MeV. For the effective electron appearance amplitude the $3\,\sigma$ allowed region is limited to $0.00048 \lesssim \sin^2(2\theta_{e \mu}) \lesssim 0.0020$.
\\\\
Furthermore, an analysis~\cite{2017dentler} including all electron (anti)neutrino disappearance results as well as NEOS and DANSS data find that even if the reactor fluxes and spectra are left free in the fit, the hint for sterile neutrinos remains at $2\,\sigma$. The allowed oscillation amplitude is lowered to $\sin^2(2\theta_{14}) \approx 0.05$ for the best fit point, while $\Delta m_{41}^2 \approx 1.8\,\mathrm{eV}^2$.

\end{document}